# A Comparison of Hashing Schemes for Address Lookup in Computer Networks

Raj Jain, Senior Member

**Abstract** – Using a trace of address references, we compared the efficiency of several different hashing functions, such as cyclic redundancy checking (CRC) polynomials, Fletcher checksum, folding of address octets using the exclusive-or operation, and bit extraction from the address. Guidelines are provided for determining the size of hash mask required to achieve a specified level of performance.

## 1 INTRODUCTION

The trend toward networks becoming larger and faster, addresses becoming larger, has impelled a need to explore alternatives for fast address recognition. This problem is actually a special case of the general problem of searching through a large data base and finding the information associated with a given key. For example, Datalink adapters on local area networks (LAN) need to recognize the multicast destination addresses of frames on the LAN. Bridges, used to interconnect two or more LANs, have to recognize the destination addresses of every frame and decide quickly whether to receive the frame for forwarding. Routers in wide area networks have to look through a large forwarding database to decide the output link for a given destination address. Name servers have the ultimate responsibility for associating names to characteristics. In these applications, hashing algorithm can be used to search through a very large information base in constant time. We also investigated caching as a possible solution but found that caching could be harmful in some cases in the sense that the performance would be better without caching [4].

To compare various hashing strategies, we used a trace of destination addresses observed on an Ethernet LAN in use. The trace consisted of 2.046 million frames observed over a period of 1.09 hours. A total of 495 distinct station addresses were observed, of which 296 were seen in the destination field.

## 2 Hashing: Concepts

Webster's dictionary defines the word 'hash' as a verb *"to chop (as meat and potatoes) into small pieces"*. Strange as it may sound, this is correct. Basically, hashing allows us to chop up a big table into several small subtables so that we can quickly find the information once we have determined the subtable to search for. This determination is made using a mathematical function, which maps the given key to hash cell $i$, as shown in Figure 1. The cell $i$ could then point us to the subtable of size $n_i$. Given a trace of $R$ frames with $N$ distinct addresses and a hash table of $M$ cells, the goal is to minimize the average number of lookups required per frame.

If we perform a regular binary search through all $N$ addresses, we need to perform $1 + \log_2(N)$ or $\log_2(2N)$ lookups per frame. Given an address that hashes to $i$th cell, we have to search through a subtable of $n_i$ entries, which requires only $\log_2(2n_i)$ lookups. The total number of lookups saved is:

$$\text{Number of lookups saved} = \sum_i r_i [\log_2(2N) - \log_2(2n_i)]$$

Here, $r_i$ is the number of frames that hash to the $i$th cell $\sum r_i = R$. The net saving per frame is:

$$\text{Lookups saved per frame} = \sum_i -\frac{r_i}{R} \log_2(\frac{n_i}{N})$$
$$= \sum_i -q_i \log_2(p_i) \quad (1)$$

Here, $q_i = r_i/R$ denotes the fraction of frames that hash to $i$th cell, and $p_i = n_i/N$ is the fraction of addresses that hash to $i$th cell. The goal of a hashing function is to maximize the quantity $\sum -q_i \log_2(p_i)$. Notice that $p_i$ and $q_i$ are not related. In the special case of all addresses being equally likely to be referenced, $q_i$ is equal to $p_i$ and the expression $\sum -p_i \log_2(p_i)$ would be called the **entropy** of the hashing function. It is because of this similarity that we will call the quantity $\sum -q_i \log_2(p_i)$ the entropy or **information** content of the hashing function. It is measured in units of 'bits.'

## 3 Hashing Using Address Bits

The simplest hashing method is to use a certain number of bits, say $m$, from the address. For example, we could hash using bits $i, i+1, \ldots, i+m-1$ of the address to $2^m$ subtables. The number of lookups saved, as we saw in the last section, is equal to the information entropy of the bits. For our trace, this is plotted in Figure 2. Eight curves corresponding to $m$ consecutive bits with $m=1, 2, \ldots, 8$ are plotted. From Figure 2, we observe that:

1. All 8-bit sequences between bits 0 and 24 have less than two bits of information.

2. The bits 32 through 39, in general, have a high information content.

The first observation is not surprising considering that the first three octets of the IEEE 802 addresses used on IEEE 802 LANs are assigned by the IEEE and, thus, most stations have the same first three octets. The first two bits corresponding to the global/local assignment and multicast/unicast addresses may be different in different addresses. Given these two bits, other bits can be easily predicted. In multivendor environments the first 3 octets may have a little more information. However, in general, these bits are not good for hashing purposes.

The second observation says that the fifth octet of the address has the highest information *in our environment*. This observation, if applicable, leads to the following types of conclusions:

1. *Use the fifth octet as the hashing function.* The bits in this octet would provide a maximum savings in the number of lookups.

2. *When comparing two addresses, compare the fifth octet first.* If the addresses are not equal, the very first comparison will fail more often than when using other octets.

3. *Use the fifth octet as the branching function at the root of a tree database.* If the addresses are stored in a tree or trie structure [5] and the address bits are used to decide the branch to be taken, using bits from this octet would provide maximum discrimination.

4. *Use the fifth octet as the load balancing function for different paths.* Given several alternative paths to a set of destinations, one way to balance the load on different paths is to decide the path based on a few bits of the address. This eliminates out-of-order arrivals since all frames going to a particular destination follow a single path and load balancing is achieved by different destinations using different paths.

It should be obvious that the fifth octet may not be the most informative octet for all environments. Nonetheless, the above recommendations are useful providing that one uses the appropriate, most informative octet instead.

# 4 Hashing Using CRC

Another hashing function, already used in a few adapters, is to use bits from the cyclic redundancy check (CRC) of the address. Using the 32-bit CRC polynomial used on IEEE 802 LANs [2], we computed the information content of $m$ bit sequences consisting of bits $i$ through $i+m-1$ of the CRC for $m=1, 2, \ldots, 8$, and $i=0, 1, \ldots, 31$. Here, $i=0$ represents the most significant bit of the CRC. The results are shown in Figure 3.

It is interesting to compare Figures 2 and 3. Notice the following:

1. Almost all 32 bits have a high information content very close to one bit. Thus, it does not matter which bit of the CRC we use.

2. All curves are (almost) horizontal straight lines. Thus, the information content of all bit combinations is identical. It does not matter which $m$ bits are chosen for $m=1, 2, \ldots$

3. The information content of $m$ bits is approximately $m$. This means that CRC provides an almost optimal hashing function.

We repeated the analysis with several other 8-bit and 16-bit CRC polynomials. In general, we found that if a polynomial provides a good CRC, it can serve as an excellent hash function.

# 5 Hashing Using Fletcher Checksum

This checksum is used in the ISO/OSI transport since it is so easy to compute in software. Given an $n$ octet message $b[1]\ldots b[n]$, a two-octet checksum $C[0]$ and $C[1]$ is computed as follows:

```
C[0] = 0; C[1] =0;
For i = 1 step 1 until n do
  C[0] = C[0] + b[i];
  C[1] = C[1] + C[0];
EndFor;
```

The information in $m$ consecutive bits of address checksums is shown in Figure 4. This also is a good hashing function.

## 6 Hashing Using Another Checksum

Another popular checksum algorithm used to guard against memory errors in network address databases is [2]:

$$C = \text{Mod}\left(2^8(4b[1]+2b[3]+b[5]) + (4b[2]+2b[4]+b[6]), 2^{16}-1\right)$$

Here, $b[i]$ is the $i$th octet of the address and $C$ is the 16-bit checksum. Since we are not aware of its name, we will call it the 'mod-checksum.' The information content of the bits of this checksum are shown in Figure 5. Notice that the mod-checksum is not as good a hashing function as the Fletcher checksum even though it is more complex to compute.

## 7 Hashing Using XOR Folding

The final alternative that we investigated is that of the straightforward exclusive-or operation on the six octets of the address to produce 8 bits.

$$C = b[1] \oplus b[2] \oplus b[3] \oplus b[4] \oplus b[5] \oplus b[6]$$

The information content of the bits in the 'XOR-fold' so obtained is shown in Figure 6. To our surprise, this function, which is so simple to implement, is an excellent hashing function.

## 8 Mask Size for an Address Filter

In this section, we briefly address the problem of finding the size of the hash mask required to get a desired level of performance. We assume that the filter consists of a simple $M \times 1$ bit mask, that is, each hash table cell is only one bit wide. A hash function is used to map the address to an index value $i$ in the range 0 through $M-1$, and if the $i$th bit in the mask is set, the frame is accepted for further processing; otherwise, the frame is rejected. This is how hashing is used in several commercial media access controller (MAC) chips. Such a hash filter is a perfect rejection filter in the sense that if the mask bit is zero, we can be certain that the address is not wanted, and there is no need to search the address table.

The performance of the filter is measured by the probability of an unwanted address being rejected by the filter. We call this probability the **unwanted-rejection rate**. If we assume that all addresses are equally likely to be seen and that all mask cells are equally likely to be referred, then using the procedure described in [3], we can compute the unwanted-rejection rate as shown in Figure 7. In the figure, the number of addresses $k$ that a station may want to receive is plotted along the horizontal axis, and the probability of rejecting an unwanted frame is plotted along the vertical axis. Eight curves corresponding to mask size $M = 2, 4, 8, \ldots, 128, 512$ bits are shown. From figure 7, we observe the following:

1. There is some filtering even with small masks. For example, if one wants to receive 10 addresses, an 8-bit mask is expected to reject 26% of the unwanted frames without further searching. Although this rate is low, the point is that it is non-zero even though the mask size is less than the number of addresses.

2. In general, it is better to have as large a mask as possible. For example, if one wants to receive 10 addresses with a 512-bit mask, 98% of the unwanted frames will be rejected without further searching.

3. If the mask size is large compared to the number of addresses to receive, that is, if $M \gg k$, the curves are linear and the unwanted-rejection rate is approximately $1 - k/M$

The last observation is helpful in deciding the mask size. Thus, if one wants to reject 80% of the unwanted frames, the mask size should be 5 times the number of addresses desired.

## 9 SUMMARY

We showed that the number of lookups saved using hashing is equal to the information content of the bits of the hashed value and compared several hashing functions. First, simple bit extraction from the address itself provides a hashing function that is easy to implement in hardware as well as software. Second, bits extracted from the CRC of the address can be used as a hashing function that is easy to implement in hardware. Third, bits extracted from the Fletcher checksum can be used as a hashing function that is easy to implement in software. Finally, exclusive-or folding of the address octets provides another hashing alternative that is easy to implement both in software as well as hardware.

We concluded that CRC polynomials are excellent hashing functions. Fletcher's checksum and folding are also good hashing functions. The mod-checksum, which is more complex to compute than Fletcher's checksum, is not as good as the latter. Although bit extraction is not as good as other alternatives, it is the simplest. The choice between bit extraction and other alternatives is basically that of computing vs storage. If we can use excess memory, bit extraction with a few more bits may provide the same information as the checksum or folding with a few less bits.

We pointed out that for a station wanting to receive $k$ addresses, the probability of rejecting unwanted frames using a simple $M \times 1$ bit mask is $1 - k/M$. This allows us to decide the mask size required for a desired level of performance.

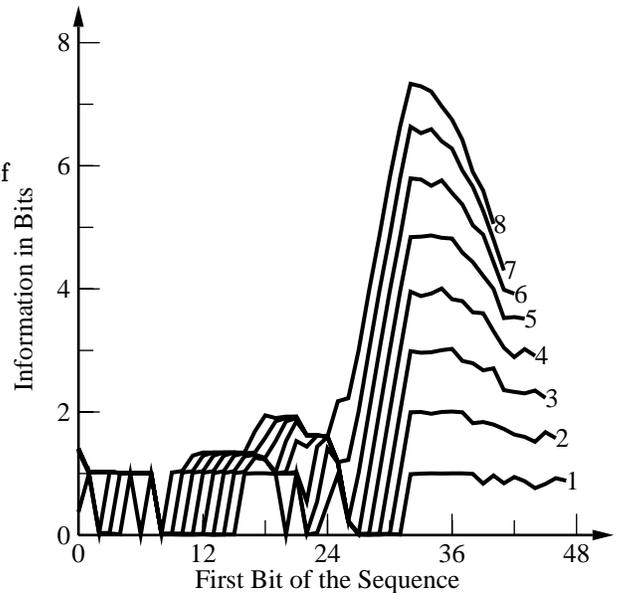

Figure 2: Information in address bits.

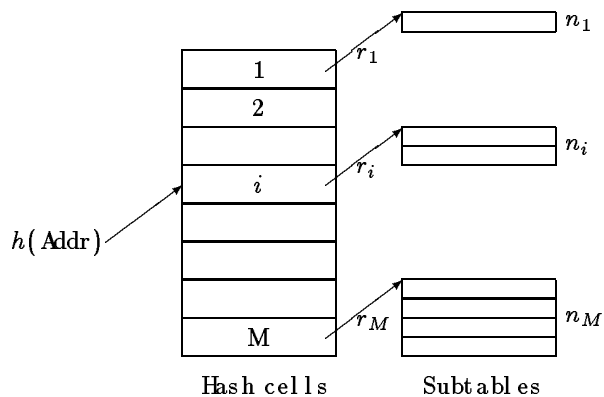

Figure 1: Hashing concepts.

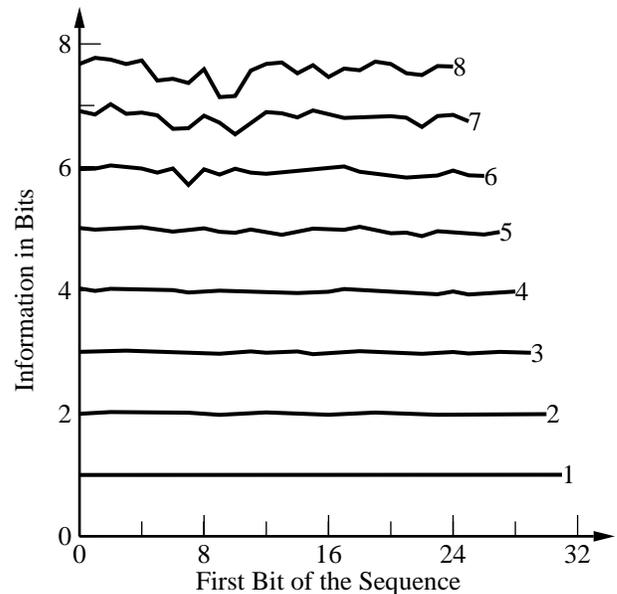

Figure 3: Information in CRC bits.

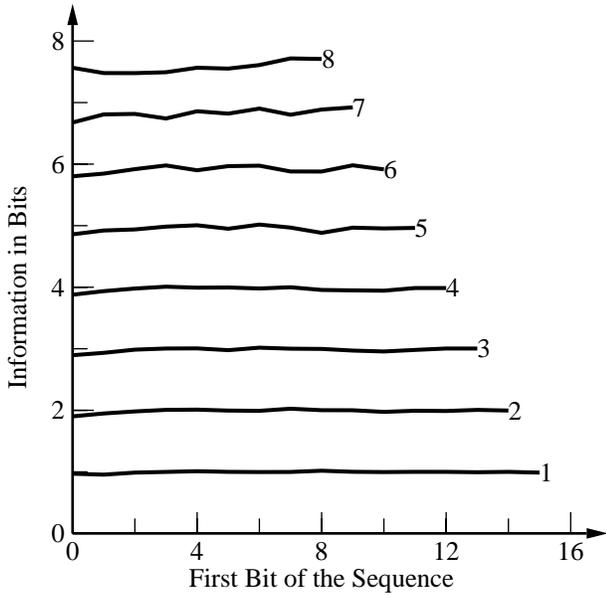

Figure 4: Information in Fletcher checksum bits.

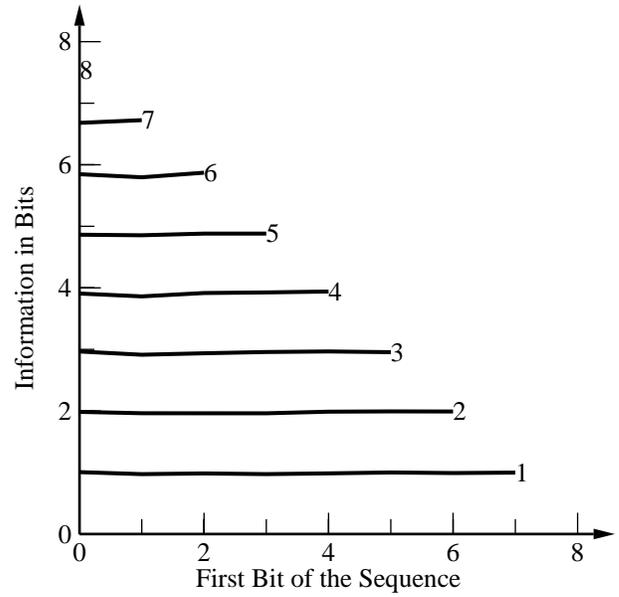

Figure 6: Information in XOR-fold bits.

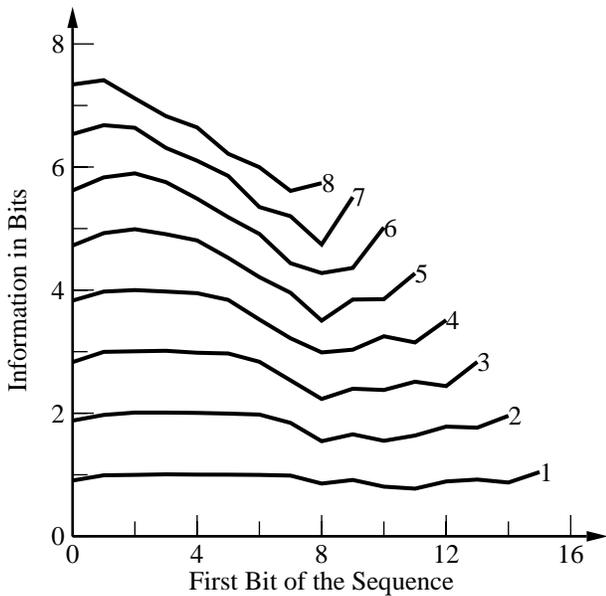

Figure 5: Information in mod-checksum bits.

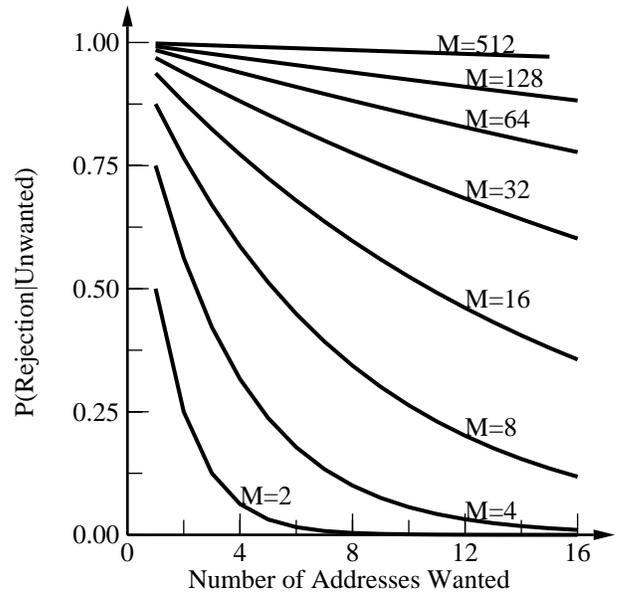

Figure 7: Probability of rejecting unwanted frames as a function of number of address wanted and the mask size $M$